\documentclass[a4paper]{jpconf}
\usepackage{graphicx}
\begin{document}

\title{Study of the $N=50$ major shell effect close to $^{78}$Ni :
First evidence of a weak coupling structure in $^{83}_{32}$Ge$_{51}$ and three-proton configuration states in $^{81}_{31}$Ga$_{50}$
}

\author{D. Verney$^1$, F. Ibrahim$^1$, O. Perru$^1$, O. Bajeat$^1$, C. Bourgeois$^1$, 
M. Ducourtieux$^1$, C. Donzaud$^1$, S. Essabaa$^1$, S. Gal\`es$^{2}$, L. Gaudefroy$^{2}$,
D. Guillemaud-Mueller$^1$, F. Hammache$^1$, C. Lau$^1$, H. Lefort$^1$, 
F. Le Blanc$^1$, A.C. Mueller$^1$, F. Pougheon$^1$, B. Roussi\`ere$^1$,
J. Sauvage$^1$ and O. Sorlin$^{2}$}

\address{$^1$ Institut de Physique Nucl\'eaire, IN2P3-CNRS/Univ. Paris Sud-XI, 91406 Orsay, France}
\address{$^2$ GANIL, Bld Henri Becquerel, 14076 Caen Cedex 5, France}
\ead{verney@ipno.in2p3.fr}

\begin{abstract}
New levels were attributed to $^{81}_{31}$Ga$_{50}$ and $^{83}_{32}$Ge$_{51}$ which were fed by the $\beta$-decay of their respective mother nuclei $^{81}_{30}$Zn$_{51}$ and $^{83}_{31}$Ga$_{52}$ produced by fission at the "PARRNe" ISOL set-up installed at the Tandem accelerator of the Institut de Physique Nucl\'eaire, Orsay. We show that the low energy structure of $^{81}_{31}$Ga$_{50}$ and $^{83}_{32}$Ge$_{51}$ can easily be explained within the natural hypothesis of a strong energy gap at $N=50$ and a doubly-magic character for $^{78}$Ni. 
\end{abstract}

\section{Experimental procedure and results}
\begin{figure}[h]
\includegraphics[width=18pc]{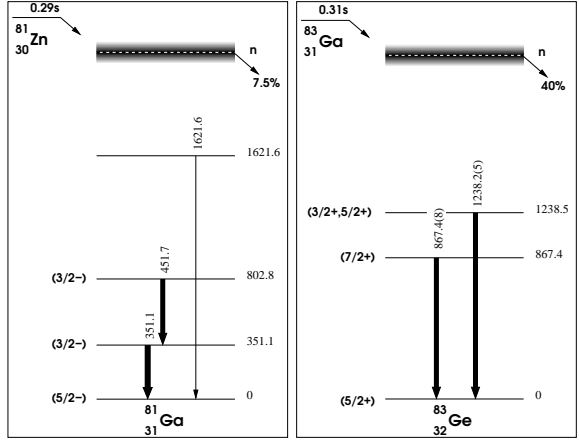}\hspace{2pc}%
\begin{minipage}[b]{16pc}\caption{\label{schem_niv} Left part : proposed level scheme for $^{81}$Ga from the present work. For the discussion on the proposed spin assignments see text (the values of $T_{\frac{1}{2}}$ and $P_n$ are taken from literature \cite{A=81}). Right part, same as left part for $^{83}$Ge (the values of $T_{\frac{1}{2}}$ and $P_n$ are taken from literature \cite{A=83}).}
\end{minipage}
\end{figure}

The sources of $^{81}$Zn ($T_\frac{1}{2}=290\pm 50$ ms) and $^{83}$Ga ($T_\frac{1}{2}=310\pm 10$ ms) were obtained at the PARRNe mass-separator facility operating on-line at the 15MV MP-Tandem at the Institut de Physique Nucl\'eaire, Orsay. In these two experiments, a $^{nat}$U target of mass $\approx 75$ g made of a series of UC$_x$ pellets heated at $2000^{\circ}\mbox{C}$ was associated with a hot plasma ion-source of the ISOLDE MK5 type and exposed to the neutron flux generated by the reaction of the 26 MeV deuteron beam delivered by the Tandem. 
Details on the detection setup can be found in Ref. \cite{perru}. The results of these experiments are summarized on Fig. \ref{schem_niv}.

\section{Structure of $^{83}$Ge}
Concerning $^{83}$Ge,
the nature of the states we observed can be inferred from a careful consideration on the systematics of low-lying positive parity states of the $N=51$ odd-nuclei. Most of these states originate from the neutron single-particle orbitals $2d_\frac{5}{2}$, $3s_\frac{1}{2}$, $2d_\frac{3}{2}$ and $1g_\frac{7}{2}$ situated above the $N=50$ closed shell and their coupling to the $2^+_1$ quadrupole first excited state of the underlying semi-magic even-even cores. In a situation of weak coupling, the energy splitting of the multiplet is governed by a simple $6-j$ symbol $\left\{ \begin{array}{ccc}
J_c & j & J \\
j & J_c & k
\end{array}
\right\}$ 
where $J_c$ stands for the core angular momentum, $j$ that for the odd particle and $J$ is the total angular momentum \cite{deshalit}. Assuming as in \cite{auerbach} a quadrupole residual interaction between the core and the particle then we take $k=2$ leading to a relative order $\frac{7}{2}^+$, $\frac{5}{2}^+$, a doublet $\frac{3}{2}^+ / \frac{9}{2}^+$ and the  $\frac{1}{2}^+$ pushed up. We propose the state at 867 keV as the $\frac{7}{2}^+$ lowest-energy member of the $2^+_1 \otimes \nu 2d_{\frac{5}{2}}$ multiplet and the state at 1238 keV as the $\frac{3}{2}^+$ or the $\frac{5}{2}^+$ member of the multiplet. The whole discussion on this spin attribution can be found in Ref. \cite{perru}.
\section{Structure of $^{81}$Ga}
As for $^{81}$Ga, our experimental results were compared to shell model calculations considering $^{78}$Ni as a core and limiting the valence space to protons. Following Ref. \cite{ji} the energy ordering for the proton sp orbitals in this valence space is $1f_{\frac{5}{2}}$ followed by $2p_{\frac{3}{2}}$. As a first guess, we assumed that the ground and the 802.8 keV states were members of the $\pi(1f_{\frac{5}{2}})^3$ configuration. For such a configuration, there exists a closed formula (by Talmi) giving the energy sequence of its members as a function of 4 parameters. We determined these parameters using the evaluated mass of $^{78}$Ni \cite{audi} as reference and the 802.8 keV value as one of the inputs. This allowed us to propose a new set of TBME $\left\langle 1f_{5/2} 1f_{5/2} \right\|V_{12}\left\| 1f_{5/2} 1f_{5/2}\right\rangle_{J=0,2,4\ T=1}$ \cite{verney}. The most remarkable difference as respect to the sets established in \cite{ji} and \cite{liset} is a strongly increased value for the term associated with the pairing $(J=0,T=1)$. The general agreement between calculated and experimental spectra of the $N=50$ odd-nuclei is improved in an unexpected way. It allows us to identify the ground state of $^{81}$Ga as the $J=5/2, v=1$ member of the $\pi(1f_{\frac{5}{2}})^3$ configuration and the 802.8 keV state as its $J=3/2, v=3$ member and, incidentally, to confirm our first guess. The state at 351.1 keV does not belong to the 3-particle configuration but has rather $J=3/2, v=1$ corresponding to a $2p_{\frac{3}{2}}$ quasi-particle like nature.


\end{document}